\documentclass{moriond}
\usepackage{amssymb}

\def\Journal#1#2#3#4{{#1} {\bf #2}, #3 (#4)}

\def\NPB{{\em Nucl. Phys.} B}

\def\PRD{{\em Phys. Rev.} D}

\def\be{\begin{equation}}
\def\ee{\end{equation}}
\def\bea{\begin{eqnarray}}
\def\eea{\end{eqnarray}}

\newcommand{\Photo}{}

\newcommand{\real}[1]{\ensuremath{\mathbb{R}^{#1}}}

\begin{document}
\vspace*{4cm}
\title{SPATIAL GEOMETRY OF THE LARGE-SCALE UNIVERSE: THE ROLE OF QUANTUM GRAVITY, DARK ENERGY AND OTHER UNKNOWNS}

\author{ M. HOLMAN }

\address{Utrecht University, \\
Heidelberglaan 8, 3584 CS, Utrecht, Netherlands}

\maketitle\abstracts{
It has been known for some time that the usual inference drawn from the observed near-flatness of the large-scale Universe - namely the existence of a cosmological ``flatness problem'',
which is then taken as a partial, but key motivation for assuming the existence of an ultra-short, inflationary expansion of the very early Universe - is in itself deeply problematic.
The present contribution consolidates the earlier results regarding the absence of a cosmological flatness problem of the sort that could potentially be resolved by inflation, by clearing
up some common misunderstandings and by presenting some arguments in more detail.}

\section{Introduction}

\subsection{Measure of Cosmological Curvature}\label{subsec:prod}

Within the class of \emph{Friedmann-Lema\^itre-Robertson-Walker} (FLRW) solutions to Einstein's equation, proximity of the Universe's large-scale spatial geometry to flatness 
is conventionally captured in terms of a continuous parameter, $\Omega \equiv \Omega_m + \Omega_{\Lambda}$, where $\Omega_m \equiv \rho / \rho_c \equiv 8   \pi   \rho / 3H^2$, 
$\Omega_{\Lambda} \equiv \Lambda / 3H^2$, $\rho$ and $\Lambda$ respectively denote the density of matter and a cosmological constant and $H \equiv \dot{a} / a$ denotes the Hubble parameter
associated with the scale factor $a$. FLRW dynamics is completely specified by the Friedmann equation 
\begin{equation}\label{Friedmanneq2}
\Omega \; = \; 1 \: + \: \frac{k}{\dot{a}^2} \; = \; 1 \: + \: \frac{k}{H^2 a^2} 
\end{equation}
where $k$ can take three possible values $k=-1$, $k=0$, $k=+1$, depending on whether spatial geometry is respectively hyperbolic, flat or elliptic.
Observations indicate that $\Omega \simeq 1$ in the present epoch and the \emph{flatness problem} of FLRW cosmologies, broadly construed, is the alleged ``improbability'' 
of this basic empirical fact
\footnote{Homogeneous cosmologies (i.e., Bianchi models) obey
a \emph{generalized} Friedmann equation
\begin{equation}\label{genFriedmann}
\Omega \: + \: \Sigma \: + \: K \; = \; 1 
\end{equation}
with $\Omega$ defined similar as in the isotropic case, $K$ a spatial curvature parameter (defined in terms of scalar three-curvature, $^{(3)}R$) and
$\Sigma$ a shear-parameter [1]. Proximity of $\Omega$ to $1$ now does not necessarily entail near-vanishing spatial curvature and the flatness problem in the way it is conventionally
characterized is therefore restricted to FLRW cosmologies (cosmological precision data of course tightly constrain $\Sigma$, justifying the FLRW approximation).}.

\subsection{Distinguishing Issues}

Closer inspection of the relevant literature however reveals that ``the'' cosmological flatness problem raises at least three distinct potential issues [2].
A first broad categorization of these can be made according to whether a particular formulation of the problem depends on the dynamical nature of $\Omega$ or not.
If it does, two basic varieties are found to exist (which are often conflated in practice, however). According to the \emph{fine-tuning argument}, the observed value for $\Omega$
at the present time implies that it had to be \emph{extremely} close to $1$ at very early times and (unless $k=0$) this is deemed highly improbable.
On the other hand, if it is accepted that $\Omega$ was indeed extremely close to $1$ at very early times, according to the \emph{instability argument}, it would be
highly improbable that $\Omega$ is still close to $1$ today. 
Versions of the flatness problem in which the time-dependence of $\Omega$ plays no effective role are essentially variations on the theme of ``anthropic coincidences''  (i.e., 
why physical parameters, such as Newton's constant, $G_{\mbox{\scriptsize N}}$, or the fine-structure constant, $\alpha$, take the particular values they do) and will be discarded
in what follows.

\subsection{Fine-tuning}

The fine-tuning argument is often simply \emph{identified} with the flatness problem -- especially within the context of inflationary models -- despite the fact that its problematic nature
has been known for quite some time (see Ref. [2] and references therein).
The key point here is that the ``fine-tuning'' is \emph{inherent} to all cosmologically relevant FLRW solutions 
-- in particular, it also occurs for solutions for which $\Omega$ is \emph{not} $\simeq 1$ after some 13.8 Gyr.
Put differently, the ``fine-tuning'' of $\Omega$ is an intrinsic part of singular FLRW models, since any such model starts with $\Omega = 1$ exactly.
This contradicts the assumption of a flat, non-divergent probability distribution of initial values of $\Omega$ around  $\Omega = 1$.
It is this (non-warranted) assumption what creates the (fine-tuning version of the) flatness problem.

\subsection{Instability}\label{instable}

\begin{figure}
\begin{minipage}{0.33\linewidth}
\centerline{\includegraphics[width=1\linewidth]{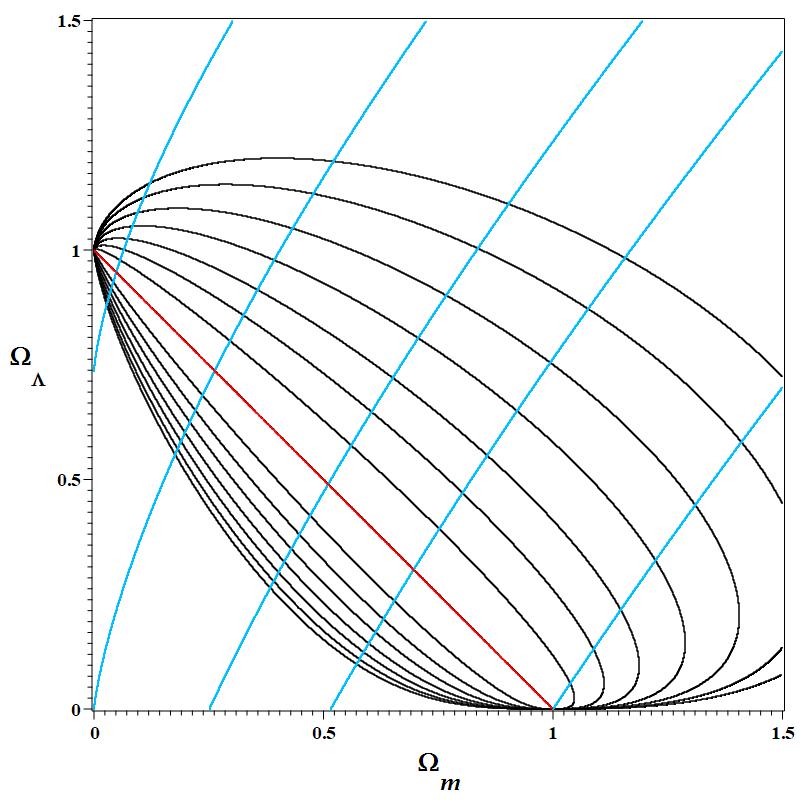}}
\end{minipage}
\caption[]{\small Level sets of $\alpha$ (in black) in the $\Omega_m \geq 0 , \quad \Omega_{\Lambda} \geq 0$ region of the phase plane, approaching the delimiter
		corresponding to a flat model (in red) from below ($k=-1$) and above ($k=+1$), as $|\alpha| \longrightarrow \infty$ (depicted are level sets corresponding
		to the specific values $|\alpha|=6,8,12,20,40,100,500$). Different universe ages (in blue) progress upward from
		the lower right to the upper left and all models evolve from $(1,0)$ (Big Bang) into $(0,1)$ (asymptotic de Sitter phase).
		Earlier observational constraints from CMB- and SN IA-data put  $|\alpha| \gtrsim 500$. Similar current constraints from Planck  
		(in particular CMB TT/EE power spectra and BAO constraints) increase this lower bound by a factor of at least $6,000$ (the level sets corresponding
		to this value of $|\alpha|$ would be indistinguishable from the flat model on the scale depicted).}
\label{fig:radish}
\end{figure}

Given that $\Omega$ was extremely close to $1$ at very early times and that, \emph{at least initially}, it necessarily moves away from $1$
in any non-flat representative model, it may seem plausible at first glance that one ``should'' not expect $\Omega$ to \emph{still} be ``close'' to $1$ today, 
i.e., after some fourteen billion odd years of cosmological evolution. 
Such expectations are easily shown to be incorrect through simple examples [2], which also provide a qualitative account of the following more rigorous argument due to Lake [3].
A fully equivalent specification of FLRW dynamics in the case of dust ($w=0$) is given by
\begin{eqnarray}
\Omega_m'	       & = & (\Omega_m - 2 \Omega_{\Lambda} - 1)\Omega_m 		   \label{FLRWdyn1}  \\
\Omega_{\Lambda}'      & = & (\Omega_m - 2 \Omega_{\Lambda} + 2)\Omega_{\Lambda}           \label{FLRWdyn2}
\end{eqnarray}
where $' \equiv d/d\eta = H^{-1} d/d\tau$, with $\eta \equiv \ln (a/a_0)$.
The key feature of the argument is that
\begin{equation}\label{FLRWconstant}
\alpha \; \equiv \; \mp \frac{27 \Omega^2_m \Omega_{\Lambda}}{4 \Omega_k^3} \qquad \quad \mbox{for} \; \; k = \pm 1, \quad \Omega_k \equiv -k/\dot{a}^2
\end{equation}
is a constant of the motion for (\ref{FLRWdyn1}), (\ref{FLRWdyn2}) (for any FLRW solution $\Omega_k = 1 - \Omega_m - \Omega_{\Lambda}$) and thus can be used to label 
dynamical trajectories in the phase plane (cf. Figure \ref{fig:radish}).
When this is done, \emph{typical} models are found to never develop any significant curvature, a property which moreover continues to hold in the more generic case (i.e., including different
components $w_i \neq 0$).




\section{Flatness of the Very Early Universe}\label{QG}

Although the foregoing conclusions seem transparent and decisive enough, doubts are sometimes expressed, although so far not really made explicit,
that some subtle point or other has been overlooked in the entire discussion.
For instance, it is obvious that when initial conditions are imposed at, say, the Planck scale, for the overwhelming majority of singular FLRW models,
$\Omega$ will still start ``very close'' to 1. The fact that it is less obvious how to express such a property in a mathematically precise way should not be
confused with evidence for fine-tuning. In order to appreciate this point better, momentarily consider a more general cosmological setting.

\subsection{Measure and Probability in Cosmology}

As is well known, general relativity has a Hamiltonian formulation, which formally gives rise to a canonical (Liouville) measure, $\mu$, on the corresponding phase space, $\Gamma$.
In practice, this picture is more involved because of (a) the necessity to ``truncate'' $\Gamma$, so that its dimension becomes finite ($\mu$ - which is proportional
to the top exterior product of the symplectic form with itself - otherwise being ill-defined), and (b) the presence of gauge constraints.
Adopting as the configuration manifold Wheeler's superspace of all Riemannian three-geometries, partially deals with the second issue,
while restricting attention to homogeneous models renders superspace finite-dimensional (``\emph{mini-superspace}'') and thus addresses the first issue.
Pull-back of the symplectic form to the corresponding physical phase space $\Gamma_{\mbox{\scriptsize ms}}$ in this case gives rise to the so-called 
\emph{GHS measure} $\mu_{\mbox{\tiny GHS}}$ [4].
If $\mu_{\mbox{\tiny GHS}}(\Gamma_{\mbox{\scriptsize ms}})<\infty$, it would then make sense to define probabilities as in ordinary statistical physics.
If $F$ denotes some particular observable feature (e.g., the property that $|\Omega - 1|<\epsilon$, $\epsilon \in \real{+}$) and $X_F \subseteq \Gamma_{\mbox{\scriptsize ms}}$ such that
$F$ holds, the probability to observe $F$ would be given by
\begin{equation}\label{GHSprob}
P(F) \; = \; \frac{\mu_{\mbox{\tiny GHS}} (X_F)}{\mu_{\mbox{\tiny GHS}} (\Gamma_{\mbox{\scriptsize ms}})}
\end{equation}
However, $\mu_{\mbox{\tiny GHS}}$ is not finite and probabilities for most measurable subsets 
are ambiguous (in fact, expression (\ref{GHSprob}) is well defined only if $X_F$ has finite measure (yielding $P(F)=0$) or if the complement of $X_F$ has finite measure (yielding $P(F)=1$)).
By implementing a ``regularization procedure'' these ambiguities can be removed, albeit in a highly non-unique way [5] (as a result, different authors have also obtained widely different 
estimates for the ``probability of inflation'').\\
Attempts to go beyond the GHS framework run into further difficulties.
Even if attention is restricted to near-FLRW solutions, as soon as inhomogeneous modes are included, a truncation is again necessary and this gives rise to several serious difficulties [5].
The upshot of the foregoing remarks is that, despite superficial appearances to the contrary perhaps, standard techniques do \emph{not} provide any information on 
how values of $\Omega$ are probabilistically distributed, not even in the simplest FLRW-context.
Furthermore, \emph{non-standard} techniques based on an objective Bayesian approach to probability \emph{do} give additional information in the form of divergent, improper measures
around $\Omega=1$ [6].


\subsection{Quantum Gravity Effects?}

It would make little sense (if at all) to argue that the absence of initial fine-tuning in $\Omega$ would somehow be overthrown by mysterious quantum gravity effects.
Although it is of course impossible - given the lack of agreement about the very identity of these effects - to say anything in this regard with absolute certainty, it would seem little short of a 
miracle if a future theory of quantum gravity were found to somehow generically predict $\Omega (t_{\mbox{\scriptsize Pl}})$ to take any value in some finite interval including $1$
with more or less equal probability. Even though it is certainly true that initial conditions at the classical level can be chosen such that $\Omega (t_{\mbox{\scriptsize Pl}})$
takes any prescribed value, it does not appear to widely appreciated that initial conditions for which $\Omega (t_{\mbox{\scriptsize Pl}})$ is not close to 1 would lead to fine-tuning issues themselves, i.e., for the integration constants [2].

\section{Flatness of the Large-Scale Late Universe}

\subsection{Dark Energy Effects?}

It may also be wondered whether the conclusions of subsection \ref{instable} could somehow be affected by unknown ``dark energy'' effects.
Even though all present cosmological data are perfectly consistent with - and perhaps even favoured by - the concordance model, i.e., $\Lambda = \mbox{constant}$ , 
representation of dark energy [7], \emph{if} $\Lambda$ would turn out to be dynamical, this could only play a role at late times - with dark energy slowly decaying away 
and potentially giving rise to curvature effects.
Since the effects of such dynamical $\Lambda$ would be equivalent to those of a cosmological constant up until the present age, the conclusions presented earlier
would retain their validity\footnote{A strong motivation that is sometimes given to prefer dynamical (i.e., ``dark energy'') interpretations of $\Lambda$,
is that these would avoid the so-called cosmological coincidence problem, of why $\rho$ and $\Lambda$ roughly have the same values in the present epoch.
This is essentially just the instability version of the flatness problem in disguise, however, and it was seen in subsection \ref{instable} that this version 
is perfectly amenable to treatment with $\Lambda$ constant!


}

\section{Conclusion}

A cosmological flatness problem - of either the fine-tuning or instability variety - cannot be said to exist.
Even though \emph{all present evidence points against it}, 
one could \emph{imagine} that fine-tuning could after all 
still \emph{become} an issue within a particular approach towards quantum gravity in the future.
Until then, it appears that a much more fruitful way of proceeding would consist in reshaping the evidence for both absence of fine-tuning and
stability of flatness over cosmological aeons into a selectional \emph{constraint} on a theory of quantum gravity.

\end{document}